\documentclass[prl,aps,superscriptaddress,twocolumn]{revtex4-1} 

\def\be{\begin{equation}}
\def\ee{\end{equation}}

\usepackage{bm}
\usepackage{amsmath,epsfig}
\usepackage{mathtools}
\usepackage{amsfonts}
\usepackage{color}
\usepackage[usenames,dvipsnames]{xcolor}
\usepackage{mathrsfs}
\usepackage{soul}   

\usepackage{hyperref}

\newcommand{\bea}{\begin{eqnarray}}
\newcommand{\eea}{\end{eqnarray}}
\newcommand{\bi}{\begin{itemize}}
\newcommand{\ei}{\end{itemize}}

\newcommand{\rr}{{\bf r}}

\newcommand{\vn}{{\bf 0}}
\newcommand{\ra}{\rangle}
\newcommand{\la}{\langle}

\newcommand{\up}{\uparrow}
\newcommand{\down}{\downarrow}

\newcommand{\Gr}{\mathcal{G}}

\newcommand{\Or}{\mathcal{O}}

\newcommand{\bl}{} 
\newcommand{\ora}{} 
\newcommand{\red}{} 
\newcommand{\mage}{} 

\begin{document}

\title{
  High-order diagrammatic expansion around BCS theory
}

\author{G. Spada}
\email{gabriele.spada@unitn.it}
\affiliation{Laboratoire Kastler Brossel, \'Ecole Normale Sup\'erieure - Universit\'e PSL, CNRS, Sorbonne Universit\'e, Coll\`ege de France,  75005 Paris, France}
\affiliation{Laboratoire de Physique de l'\'Ecole Normale Sup\'erieure, ENS - Universit\'e PSL, CNRS, Sorbonne Universit\'e, Universit\'e de Paris, 75005 Paris, France}
\affiliation{Pitaevskii BEC Center, CNR-INO and Dipartimento di Fisica, Universit\`a di Trento, I-38123 Trento, Italy}

\author{R. Rossi}
\email{riccardo.rossi@cnrs.org} 
\affiliation{Center for Computational Quantum Physics,
The Flatiron Institute, New York, NY 10010, USA
}
\affiliation{Institute  of  Physics, \'Ecole  Polytechnique  F\'ed\'erale  de  Lausanne, 
  1015  Lausanne,
  Switzerland}
\affiliation{Sorbonne Université, CNRS, Laboratoire de Physique Théorique de la Matière Condensée, 75005 Paris, France}

\author{F. $\check{\rm S}$imkovic}
\affiliation{CPHT, CNRS, \'Ecole Polytechnique, Institut Polytechnique de Paris, 
91128 Palaiseau, France}
\affiliation{Coll\`ege de France, 11 place Marcelin Berthelot, 
75005 Paris, France}

\author{R. Garioud}
\affiliation{CPHT, CNRS, \'Ecole Polytechnique, Institut Polytechnique de Paris, 
91128 Palaiseau, France}
\affiliation{Coll\`ege de France, 11 place Marcelin Berthelot, 
75005 Paris, France}


\author{M. Ferrero}
\affiliation{CPHT, CNRS, \'Ecole Polytechnique, Institut Polytechnique de Paris, 
91128 Palaiseau, France}
\affiliation{Coll\`ege de France, 11 place Marcelin Berthelot, 
75005 Paris, France}

\author{K. Van Houcke}
\affiliation{Laboratoire de Physique de l'\'Ecole Normale Sup\'erieure, ENS - Universit\'e PSL, CNRS, Sorbonne Universit\'e, Universit\'e de Paris, 75005 Paris, France}

\author{F. Werner}
\email{werner@lkb.ens.fr}
\affiliation{Laboratoire Kastler Brossel, \'Ecole Normale Sup\'erieure - Universit\'e PSL, CNRS, Sorbonne Universit\'e, Coll\`ege de France,  75005 Paris, France}

\date{\today}

\begin{abstract}
  We demonstrate that
  summation of connected diagrams to high order
  starting from
  a BCS hamiltonian
  is a viable generic
  unbiased approach for strongly correlated fermions in superconducting or superfluid phases.
  For the 3D attractive Hubbard model in a strongly correlated regime,
    we observe convergence of the diagrammatic series, 
    evaluated up to 12~loops thanks to the connected determinant diagrammatic Monte~Carlo algorithm.
    Our study includes the polarized regime, where
          conventional quantum Monte Carlo methods suffer from the fermion sign problem.
Upon increasing the Zeeman field,
we observe the first-order superconducting-to-normal phase transition
at low temperature,
{\bl and a thermally activated polarization of the superconducting phase well described by quasiparticle theory.}

  
\end{abstract}
\maketitle

After the discovery of superconductivity 110~years ago~\cite{K_Onnes},
it took nearly half a century before
Bardeen, Cooper and Schrieffer provided a microscopic explanation
based on an
ansatz for the many-body ground-state
wavefunction -- a coherent state of pairs, breaking the $U(1)$ symmetry corresponding to particle number conservation~\cite{BCS_both}.
Variational minimization over this ansatz leads to the
well-known BCS mean-field theory which captures not only the ``BCS regime'' where the attractive interaction is weak, but also the ``BEC regime'' where the attractive interaction is strong, suggesting a smooth crossover from a fermionic superfluid with large Cooper pairs to a Bose-Einstein condensate of small composite bosons~\cite{Eagles,Leggett_both}.
 This BCS-BEC crossover scenario,
 confirmed experimentally in ultracold atomic gases~\cite{RevueTrentoFermions,RevueBlochDalibard, *BlochDalibardNascimbene,ZwergerBook},
 is relevant to
neutron matter~\cite{GandolfiRevueNeutrons,StrinatiUrbanReview}
and to various solid-state materials~\cite{RevueAttractiveHubb,NSR}
where $s$-wave pairing arises between opposite-spin electrons 
\cite{Capone_C60_RMP, *FeSe_BEC-BCS, *FeSeTe_Tuning_Crossover, *HerreroMirror, *Yoshihiro_Crossover}
or between an electron and a hole 
\cite{KeldyshKozlovExcitonBEC, *CombescotRevueExcitonBEC, *Fauque_graphite_BEC}.
The problem becomes even more interesting
in presence of a Zeeman field $h$,
{\it i.e.}, a chemical potential offset between $\up$ and~$\down$~fermions,
which favors a difference between $\up$ and $\down$ densities, and tends to destabilize the 
fully paired superconducting state.


A minimal theoretical formulation of the BCS-BEC crossover problem 
is 
the attractive Hubbard model on the cubic lattice, which was widely studied
at $h=0$
(and generic filling~\footnote{The attractive and repulsive model are related by a transformation that exchanges doping
  {\bl (with respect to half-filling)} and polarization.
In particular,
the attractive 
doped model is equivalent to the repulsive polarized 
model, which is also relevant to materials~\cite{GeorgesKrauthMag}. 
The
{\bl unpolarized repulsive model, equivalent to the half-filled attractive model,} is a special case where the broken symmetry is $SO(3)$,
and {\bl is the subject of a separate study~\cite{GarioudPreprint}.}})
by
different versions and extensions of
BCS mean-field theory~\cite{MicnasMF,TormaGMB} and of the T-matrix approximation~\cite{NSR,Engelbrecht_SC_Tmatrix, *Miyake2008},
dynamical mean-field theory (DMFT) in the normal \cite{GeorgesKrauthMag, JarrellAttract, *SchollwoeckAttract, *CaponeAttract, *RanderiaAttract3D, *ToschiDMFT_Attract_Norm, *BauerPG, *ImadaCivelliAttract}
and the superconducting~\cite{RanderiaDMFT, *ToschiStiffness, *BauerGap,BauerDupuis, *KogaWerner_balanced, *AokiHolsteinHubb, *SadovskiiBalanced} phase,
and the dynamical vertex approximation~\cite{ToschiDGammaAttract}.
Unbiased
studies,
based on
the auxiliary field quantum Monte Carlo~(AFQMC)~\cite{BeckAFQMC,CarlsonAFQMC,*Drut_update,*KaplanXi,*AlhassidPairing,*AlhassidC,BulgacPG_2020}
or determinant diagrammatic Monte~Carlo (DDMC)~\cite{zhenyaPRL,GoulkoPRA} methods,
are mostly restricted to
a Zeeman field $h=0$:
In the 
$h\neq 0$ regime,
these methods are
plagued by the infamous fermion sign problem~\footnote{For fermions or frustrated spins, unbiased QMC methods 
  typically
  face
  an exponential scaling of computational time with
    system size,
  which 
  limits the accessible system sizes despite tremendous efforts in various fields including condensed matter, chemistry, and lattice QCD
  \cite{RevueLattQCD, *ForcrandQCD, *Alavi_without_FN, *ScalettarDisordWithSignPb}.
  Often the sign problem 
  is  eliminated
using
  an approximate ansatz for the nodal surface~\cite{CeperleyFN_EG_1980, *ZhangKrakauer, *CeperleyReview2010,Giorgini_unb,GandolfiRevueNeutrons, *JuilletPhaseless, *ZhangReview2017, *ZhangPRX2020}
or for the entire wavefunction~\cite{CeperleyReview2010, *OhgoeHolsteinHubbard, *OhgoeCuprates, *BagrovMLsign, *CastelnovoMLsign}.
There are also efforts to develop
unbiased sign-free approaches~\cite{DrutRevueCL, *RevueThimblesAlexandru,BulgacPG_2020}.}
and
most studies resort to
the static~\cite{MicnasMF,TormaFFLO_NJP,*TormaFFLO_3D_PRL,*TrivediFFLO,*ZhangFFLO_MF_3D,TrivediRevue_SC_INS}
or
dynamical~\cite{DaoPolarSF, *KogaWerner_SocJap, *Sadovskii2018} mean-field approximations.
A very different
route is to emulate the Hubbard model with cold atoms,
although long-range order
in 3D
was not reached so far~\cite{RevueBlochDalibard, *BlochDalibardNascimbene,HuletAF, *GreinerAF, *DeMarcoBadMetal, *BakrCantedAF, *BakrTransport, *BakrARPES, *KohlAttractive, *KohlBilayer, *KohlPairCorrel, *ZwierleinMottTransport, *ZwierleinDoublonHole, *ZwierleinPairingLatt, *BlochMagPolaron, *BlochDoped2021}.
%

In this Letter, we demonstrate that unbiased accurate results in the polarized superconducting phase can be obtained from a high-order diagrammatic expansion around a BCS hamiltonian.
 By extending 
  the connected determinant (CDet) algorithm~\cite{RossiCDet} to anomalous propagators,
  we 
  go up to
  twelve-loop order
  and observe convergence of the series.
This extends the realm of controlled 
diagrammatic computations for strongly correlated fermions in the thermodynamic limit~\cite{KozikVanHouckeEPL, *KulaginPRL, *MishchenkoProkofevPRL2014, *GukelbergerPwave, *DengEmergentBCS, *SimonsHubbardBenchmark, *HuangPyro, *IgorCoulombPhonon, *IgorDirac, *kozik_pseudogap, *SimonsHydrogenChain, *CarlstromWeyl, *JohanInfiniteU, *IgorHaldane, *KunHauleEG, *SimkovicCrossover, *KimKozik2020, *KozikEntropy, *LeBlanc_suscept, *Vucicevic2021, *SchaeferFootprints, *Mishchenko_elec_ph_blockade, *WietekTriang, *LenihanTc, *LeBlanc_EG_real_freq, *SimkovicPG, VanHouckeEOS, *RossiEOS, *RossiContact, RossiCDet, RossiRDet, *SimkovicSuscept_k}
to superconducting phases.
We determine the critical Zeeman field where a first-order superconducting-to-normal phase transition takes place at low temperature, and find a significant polarization of the superconducting phase at higher temperature.
Our results deviate very substantially from the BCS mean-field predictions
and provide reliable benchmarks for optical-lattice experiments.

The Hubbard model is defined by the 
hamiltonian
\be
H = H_{\rm kin} - \sum_{\sigma= \up, \down} \mu_\sigma\, \hat{N}_\sigma + H_{\rm int}
\ee
with
$\mu_{\up/\down} = \mu \pm h$  the chemical potentials,
$H_{\rm kin} = -t \sum_{\la\rr, \rr' \ra \,\sigma} (c_{\rr \sigma}^\dagger c_{\rr' \sigma} + h.c.)$
the nearest-neighbor hopping, 
and $H_{\rm int} = U \sum_{\rr} \hat{n}_{\rr \up}\, \hat{n}_{\rr \down}$
the on-site interaction
($c_{\rr \sigma}$ are the fermion annihilation operators,
while $\hat{n}_{\rr \sigma} = c_{\rr \sigma}^\dagger c_{\rr \sigma}$ and
$\hat{N}_\sigma = \sum_\rr \hat{n}_{\rr \sigma}$ are the single-site and total particle-number-operators).

To set up a diagrammatic expansion for the infinite-size system in the superconducting phase, 
{\red where the $U(1)$ symmetry is spontaneously broken, we expand around the unperturbed quadratic hamiltonian}
\be
H_0 = H_{\rm kin} - \sum_\sigma 
\mu_{0, \sigma}\, \hat{N}_\sigma +
H^{(\Delta_0)}_{\rm pair}
\ee
containing a symmetry-breaking pairing term
\be
H^{(\Delta_0)}_{\rm pair} :=
\Delta_0 \sum_{\rr} c_{\rr \up}^\dagger c_{\rr \down}^\dagger + h.c.
\ee
The most natural choice for the free parameters $\Delta_0$ and $\mu_{0,\sigma}$
is given by the self-consistency conditions of BCS mean-field theory
\bea
\Delta_0 &=& -U \, \la \hat{\Or} \ra_{H_0}
\\
\mu_{0,\sigma} &=& \mu_\sigma - U\,\la \hat{n}_{\vn, -\sigma} \ra_{H_0}
\label{eq:mu0}
\eea
%
where $\la \hat{\Or} \ra := \la c_{\vn \up} c_{\vn \down} \ra$
is the order parameter for the superconducting phase with long-range order in the $s$-wave pairing channel.
In what follows we will denote this mean-field choice of $\Delta_0$ by $\Delta_{\rm MF}$.
We will also use other values of $\Delta_0$, 
but always keep the mean-field choice (\ref{eq:mu0}) for the unperturbed chemical potential.

We then
introduce a hamiltonian that depends on a formal parameter $\xi$,
\be
H_\xi = H_0 + \xi\ (H-H_0),
\ee
expand intensive observables in powers of $\xi$, and finally set $\xi=1$.
For the order parameter, this means
setting
\be
\Or(\xi) := \la \hat{\Or} \ra_{H_\xi} \equiv {\rm Tr}(\hat{\Or}\,e^{-\beta H_\xi} ) \, /\, {\rm Tr }\, e^{-\beta H_\xi}
\label{eq:O_xi_def}
\ee
 (with $\beta \equiv 1/T$)
and expanding
$\Or(\xi) = \sum_{N=0}^\infty \Or_N\,\xi^N$.
\\{\red In many cases,}
this series
converges at $\xi=1$; we can 
{\red then} obtain the physical order parameter
simply by evaluating the series $\sum_{N=0}^\infty \Or_N$.
{\red 
More generally, 
since $H_0$ already breaks the $U(1)$ symmetry,
it is not necessary to cross a phase transition when increasing $\xi$ from 0 to 1,
which would prevent one from {\mage obtaining the equilibrium values of} observables at the physical point $\xi=1$ by summing or resumming their Taylor expansions at $\xi=0$.}


{\it Thermodynamic limit and spontaneous symmetry breaking.}
Here it is  
{\red conceptually important}
to work directly in the thermodynamic limit~\footnote{For an analogous discussion in the contect of the broken-symmetry phase of $\phi^4$ theory, see
    \href{https://doi.org/10.1007/JHEP05(2019)047}{M. Serone, G.~Spada, and G. Villadoro, JHEP {\bf 5}, 47 (2019)}.
}.
This limit should be taken in
the definition~(\ref{eq:O_xi_def}) of
$\Or(\xi)$,
and hence 
the thermodynamic limit should be taken before summing the $\Or_N$ over~$N$.
Indeed, recall that 
in presence of spontaneous symmetry breaking,
the order parameter is defined by introducing an external symmetry-breaking field $\eta$
that couples to the order parameter,
and sending $\eta$ to zero {\it after} taking the thermodynamic limit:
\be
\Or =\ \lim_{\eta \to 0^+}\ \   \lim_{L\to\infty} \ \ \la \hat{\Or} \ra_{H^{(\eta)}, \, L}
\label{eq:O_def}
\ee
where $L$ is the linear system size and $H^{(\eta)} := H + H^{(\eta)}_{\rm pair}$.
Let us denote by $\Or_L(\xi)$ and $\Or_{N,L}$ the finite-system versions of
$\Or(\xi)$ and $\Or_N$. 
Since there is no spontaneous symmetry breaking for a finite system,
$\Or_L(\xi=1) = 0$.
What we should do instead, to obtain the order parameter defined in (\ref{eq:O_def}), is to first take the thermodynamic limit:
$\Or = \lim_{\xi\to 1^-} \lim_{L\to\infty} \Or_L(\xi)$.
This follows simply from 
the fact that
$H_\xi$ contains a symmetry-breaking field which by construction vanishes 
in the limit $\xi\to1$ where the symmetry of the physical hamiltonian is restored. Explicitly,
$H_\xi = H_{\rm kin}
- \sum_\sigma \left[ (1-\xi) \, \mu_{0,\sigma} + \xi \, \mu_\sigma \right]\,N_\sigma
+ (1-\xi)\,H_{\rm pair}^{(\Delta_0)}
+ \xi \ H_{\rm int}$,
which is equal to $H^{(\eta_{\rm eff} = (1-\xi)\,\Delta_0)}$
plus corrections that have no effect to leading order in the limit $\xi\to1$.

{\it Diagrams and CDet algorithm.}
Each coefficient $\Or_N$ is a sum of connected Feynman diagrams with $N$ vertices.
We compute these coefficients up to a maximal order $N_{\rm max}$ using the CDet algorithm 
generalized to the broken-symmetry phase.
In addition to the normal propagator lines, 
 diagrams contain anomalous propagator lines,
where particles are destroyed at both ends,
or created at both ends.
These 
anomalous propagators are the 
off-diagonal elements of the 2~by~2 propagator matrix
$\Gr_{\alpha  \alpha'}(X-X') = - \la {\rm T}\ \Psi_\alpha^\dagger(X)\,\Psi_{\alpha'}(X') \ra_{H_0}$
with the Nambu spinor notation
$(\Psi_0, \Psi_1) := (c_\up, c^\dagger_\down) $.
Here $X \equiv (\rr,\tau)$ stands for space and imaginary-time,
and ${\rm T}$ is the time-ordering operator.

        
Following the CDet approach, we express the diagrammatic series for the order parameter ($\hat{Q} := \hat{\Or}$) or for the densities ($\hat{Q} := \hat{n}_{\vn,\sigma}$) as
%
$\la \hat{Q} \ra_{H_\xi}  = Q_0 - \sum_{N=1}^\infty (\xi\,U)^N \int dX_1 \ldots dX_N\ \,{\rm cdet}_Q(X_1,\ldots,X_N)\  /\  N!$
where $\int dX := \sum_\rr \int_0^\beta d\tau$, 
and ${\rm cdet}_Q(X_1,\ldots,X_N)$ is the symmetrized
sum of all connected Feynman diagrams with internal vertex positions $X_1,\ldots,X_N$
{\bl and one external point at $X=(\vn,0)$ where the operator $\hat{Q}$ is acting.
This function cdet is evaluated efficiently, in only $3^N$ operations, which is much faster than naive summation over the 
factorial number of connected diagrams.
The trick is to recursively subtract out all disconnected diagrams
from the sum of all connected plus disconnected diagrams,
the latter being given by the determinant of
a
matrix constructed from the propagators $\mathcal{G}_{\gamma\gamma'}(X_i-X_j)$~\cite{SMbcs}.}
We will also evaluate the series for the pressure,
$P(\xi) := \ln {\rm Tr} \exp(-\beta H_\xi) / (\beta L^3)
= \sum_{N=0}^\infty P_N \xi^N$,
whose coefficients $P_N$ are given by fully closed diagrams,
and can be computed with CDet in a similar way.
We use
a recently introduced many-configuration 
Monte Carlo algorithm~\cite{MCMCMC}
to carry out
the integration over the internal vertex positions 
for all diagram orders $N\leq N_{\rm max}$ at once.


{\it Results.}
Taking the hopping $t$ 
as unit of energy,
we~set $U = -5$,
and 
$\mu = -3.38$ so that the density $n = n_{\up} + n_{\down}$ is close to $0.5$ particles per site, {\it i.e.}
quarter filling -- a standard choice of generic filling that differs from the special half-filled case. 
For $h=0$, AFQMC is sign free and provides the critical temperature curve $T_c(U)$~\cite{BeckAFQMC}:  Our choice of $U$
lies in the strongly correlated regime where the curve has a broad maximum
-- we have ${T_c(U=-5)} \approx 0.25$, which is not far from 
the maximal value 0.33,
and much larger than in
 the weak-coupling regime where $T_c$ decreases exponentially with $1/|U|$.

 \begin{figure}
             \includegraphics[width=\columnwidth,trim=0.8cm 0.8cm 1.7cm 0.5cm,clip]{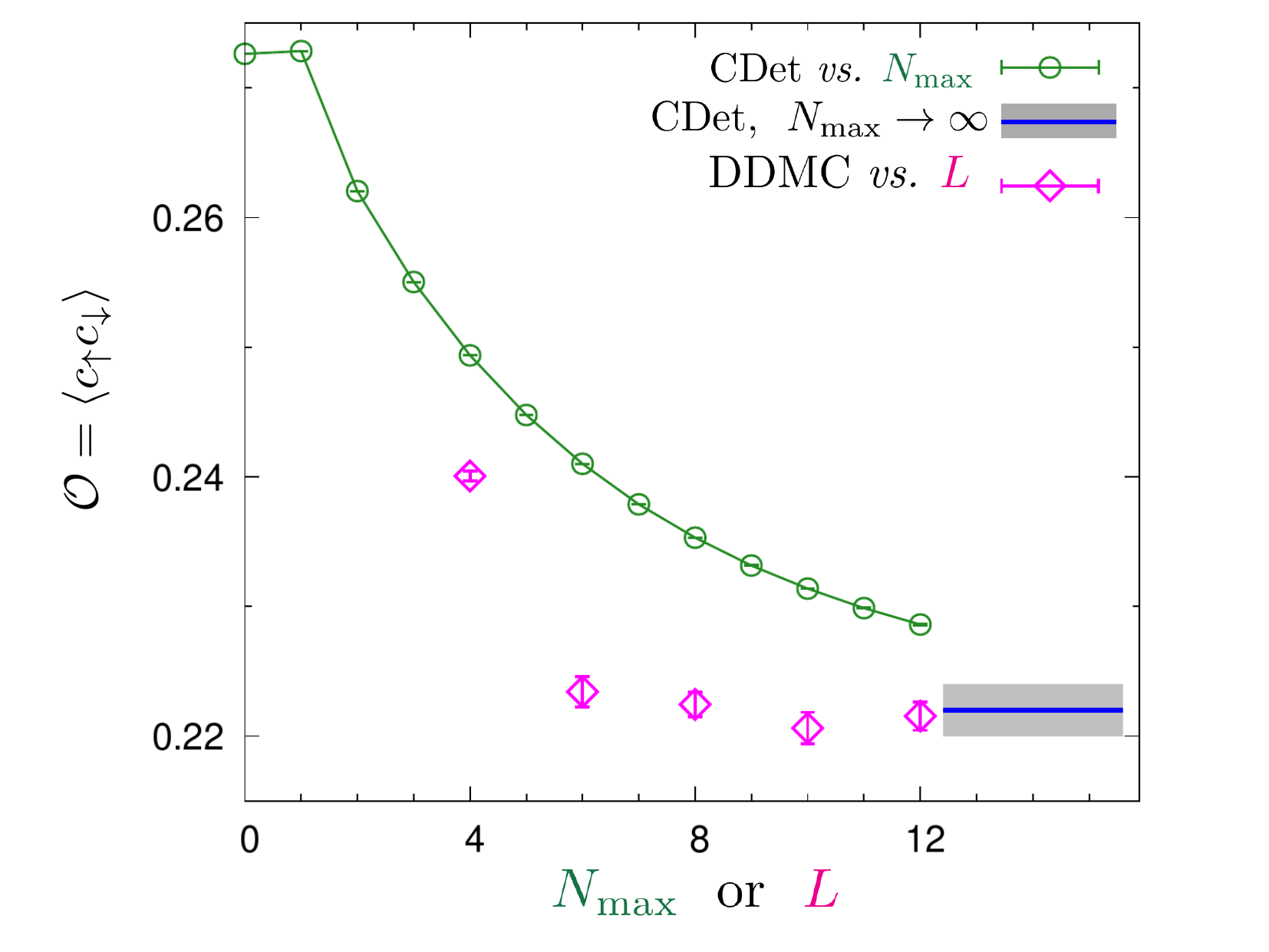}
             \caption{{\it Benchmark at zero Zeeman field:} Order parameter 
               at $T=1/8 \approx T_c/2$.
               Green circles: 
               {\mage diagrammatic expansion
               around BCS mean-field theory
                truncated at order $N_{\rm max}$.}
Blue line with grey error-band: $N_{\rm max}\to\infty$ extrapolated result.
Pink diamonds: DDMC benchmark {\it vs.} system size $L$.
\label{fig:bench}}
          \end{figure}

We start 
with a benchmark
at $h=0$. 
We compute the order parameter at $T = 1/8 \approx T_c / 2$
and compare with the DDMC method~\cite{Rubtsov2003, *Rubtsov2005,zhenyaPRL} also known as continuous-time interaction expansion in the context of impurity solvers~\cite{Rubtsov2004, *GullRevueImpurityQMC}.
Our data for the partial sum $\sum_{N=0}^{N_{\rm max}} \Or_N$ converge as a function of the truncation order $N_{\rm max}$ to a result which agrees with the DDMC benchmark, see Fig.~\ref{fig:bench}. 
Here 
we used Pad\'e approximants for the $N_{\rm max}\to\infty$  extrapolation~\cite{SimkovicCDet}. 
We used $\Delta_0=\Delta_{\rm MF}$ and checked that the extrapolated results agree 
for different choices of $\Delta_0$. 

\begin{figure}
           \includegraphics[width=\columnwidth,trim=2cm 0.6cm 0.7cm 0.33cm,clip]{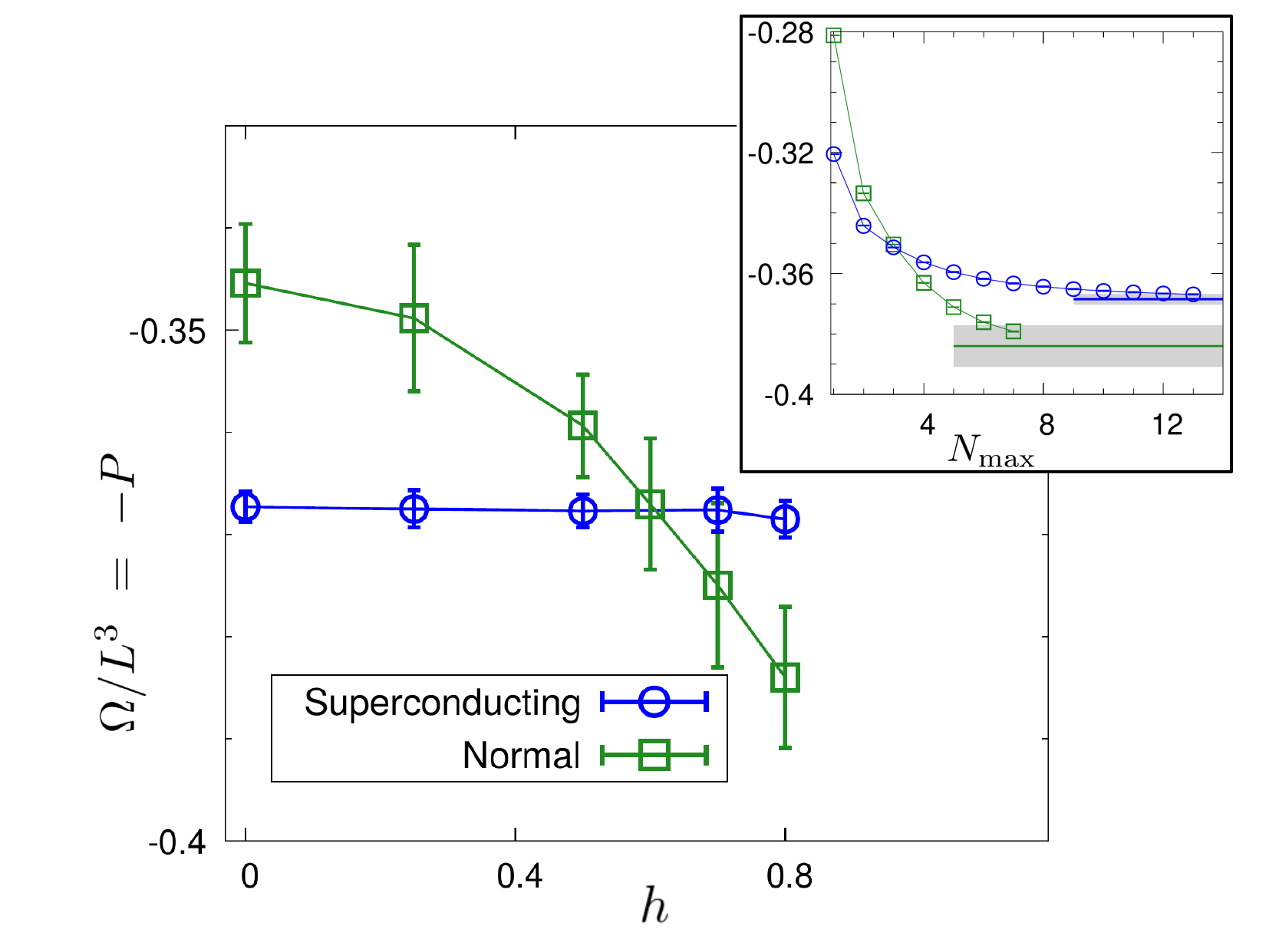}
          \caption{Grand-potential density 
            {\it vs.}
            Zeeman field,
            at ${T = 1/16} \approx T_c/4$.
            Circles: superconducting phase, obtained by expanding around BCS mean-field theory ($\Delta_0 \approx \Delta_{\rm MF}$).
            Squares: normal phase, obtained by expanding
            around the normal mean-field solution ($\Delta_0 = 0$).
            The crossing between the curves 
            signals the first-order phase transition.
            Inset: same quantity {\it vs.} truncation order $N_{\rm max}$, at $h=0.8$; horizontal lines with error bands are the
            $N_{\rm max}\to\infty$ extrapolated results also shown in the main panel.  \label{fig:P_h}}
          \end{figure}

We turn to the polarized regime $h>0$, where
conventional approaches such as
AFQMC and DDMC have a sign problem and
unbiased results are unavailable.
We start by setting the temperature to $T \approx T_c / 4$, 
increase the Zeeman field $h$, and compute the thermodynamic grand potential per unit volume, $\Omega/L^3 = -P$ with $P$ the (electronic) pressure.
We obtain the pressure 
of the superconducting phase
using again the expansion around 
the broken-symmetry mean-field solution ($\Delta_0 = \Delta_{\rm MF}$).
We also evaluate the expansion 
around the normal mean-field solution ($\Delta_0 =0$)
which yields the normal-phase pressure. 
As shown in Fig.~\ref{fig:P_h}, the two curves cross,
which indicates a first order phase transition.
The error bars are dominated by the $N_{\rm max}\to\infty$ extrapolation,
and are larger for the normal phase {\mage where} we could only evaluate the series up to order 7, instead of 12 for the superconducting phase.
We attribute this difference to the fact that the superconducting-phase propagators are gapped, and hence decay faster with position, which reduces the Monte~Carlo variance.
Within error bars, the superconducting pressure is independent of $h$, which means that the magnetization $m := n_\up - n_\down$ is zero.
This indicates that we are in the regime where
$h$ is smaller than the pairing gap $E_g$,
{\it i.e.} the Zeeman field is not large enough to overcome the energy cost
for adding 
an extra ``unpaired'' fermion,
and the magnetization is exponentially suppressed at low temperature, $m \lesssim e^{-\beta(E_g-h)}$. 
So the pairing gap essentially prevents
the superconducting phase from polarizing, until a first-order phase transition occurs when the polarized normal phase becomes energetically favorable.
Ultracold atom experiments~\cite{ShinPhaseDiag, *SylEOS, *navon2010Ground} and fixed-node Monte~Carlo calculations~\cite{Giorgini_unb} in continuous space are consistent with this scenario. 
This is also what is predicted by BCS mean-field theory~\cite{Sarma,MicnasMF}
albeit with a critical field 
nearly twice larger than our unbiased result 
$h_c = 0.61(12)$.

For $h>h_c$ the superconducting phase is metastable.
We have checked that the order parameter is still non-zero at $h=0.8$.
In this regime the convergence of the series $\sum \Or_N$ is slower and the extrapolation becomes less stable. Therefore, instead of computing the order parameter directly, we extracted it from the response to a small symmetry-breaking field:
$2\,\Or = dP^{(\eta)}/d\eta|_{\eta=0^+}$, where $P^{(\eta)}$ is the pressure in presence of the field $\eta$ ({\it i.e.} for the hamiltonian $H^{(\eta)}$), whose expansion can be extrapolated reliably.
As always, the notion of metastable phase has to be taken with a grain of salt: It is only well defined asymptotically close to the first-order transition point, where the energy barrier for nucleating the stable phase inside the metastable phase diverges.
Accordingly, the diagrammatic expansion must actually diverge, but  as long as we are not too deep in the metastable regime,
this divergence is slow and only visible at very large orders. 
Similarly, the normal phase is metastable for $h<h_c$,
and we are able to follow it all the way to $h=0$ without encountering the divergence of the series within the 7 orders that are accessible to us. 



\begin{figure}
\includegraphics[width=\columnwidth,trim=0.5cm 0.7cm 0 0,clip]{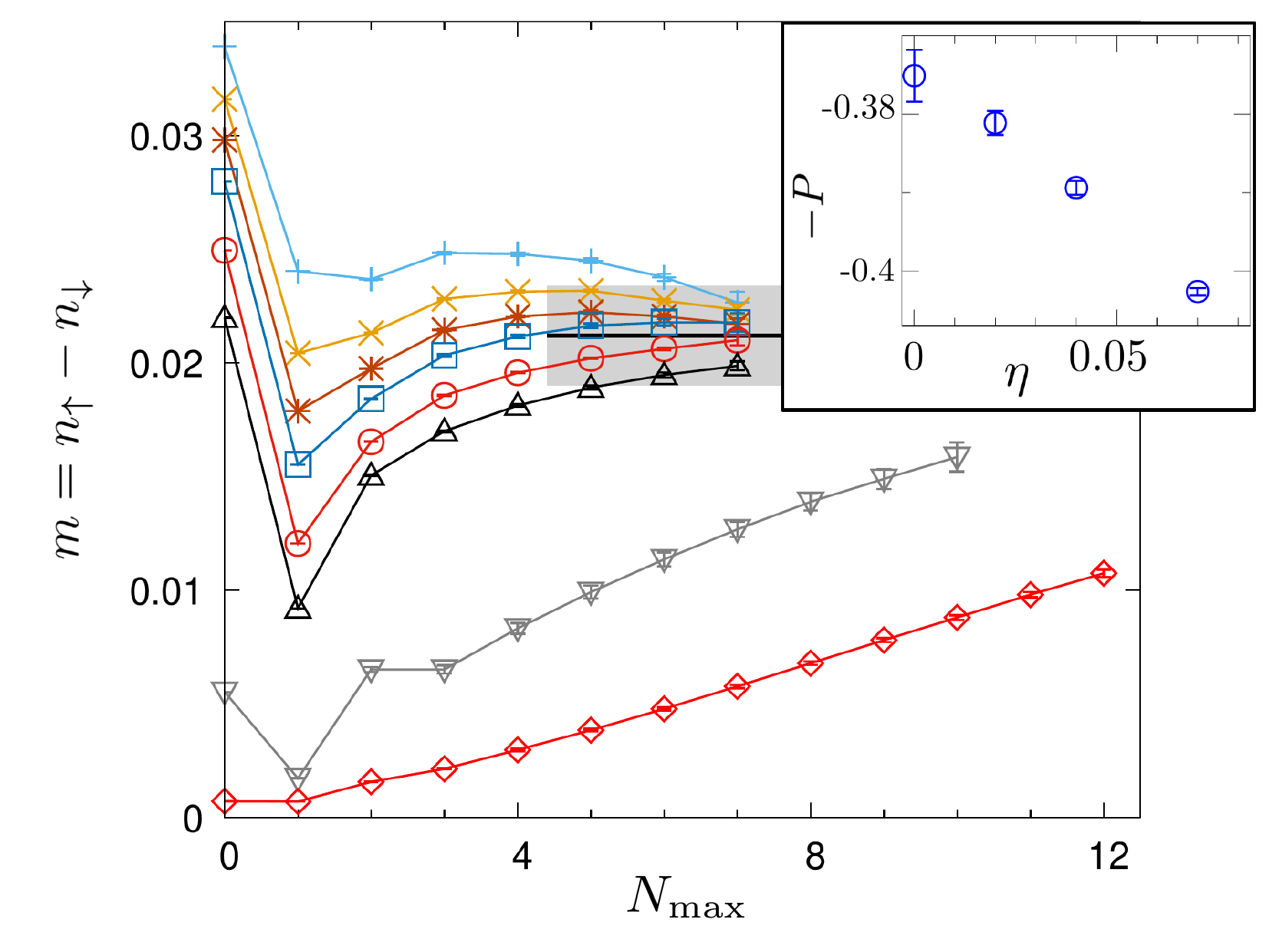}
\caption{Magnetization {\it vs.} maximal expansion order,
at~${T=0.19}\approx 3\,T_c/4$ and $h=0.35$,
for different choices of the unperturbed pairing field (from bottom to top: $\Delta_0=1.357 \approx \Delta_{\rm MF},\ \Delta_0=0.9,\, 0.5,\, 0.45,\, 0.4,\, 0.37,\, 0.34$, and 0.3) from which we obtain $m = 0.021(2)$ (horizontal line with error-band).
The BCS mean-field result (the value of the bottom curve at $N_{\rm max}=0$) is 30~times smaller.
Inset: Pressure {\it vs.} external symmetry-breaking field $\eta$, whose slope at the origin, and hence the order parameter, is non-zero.
  \label{fig:m}}
          \end{figure}


{\bl We turn to higher temperature, where we can resolve the polarization of the superconducting phase, although it remains suppressed by the pairing gap.}
At $T \approx 3\,T_c/4$ and $h=0.35$, we find a magnetization 
{\mage $m = 0.021(2)$,}
which corresponds to a polarization $(n_\up - n_\down)/(n_\up+n_\down)$ of 4\%.
This is 
30 times larger than the BCS mean-field prediction.
Therefore, BCS mean-field is not a good starting point for the expansion in this case,
and we had to tune $\Delta_0$ away from $\Delta_{\rm MF}$ in order to obtain 
convergence of the partial sums
within accessible orders, see Fig.~\ref{fig:m}
{\ora (we used the Fastest Apparent Convergence principle to produce the final value and error bar).} 
Furthermore we can again check that the order parameter is 
non-zero by computing $P$ {\it vs.} external field~$\eta$, see inset of Fig.~\ref{fig:m}.
We thus observe a polarized superconducting state. 
This state 
is possibly metastable, since its pressure (at $\eta=0$) does not differ from the one of the normal phase within our error bars. 

{\bl To probe 
the nature of this polarized superconducting state,
we repeat the computation of the magnetization for different values of $h$ at fixed $T$,
and 
fit our data with the expression
$m(h,T) \simeq n_{\rm qp}(T)\,{\rm sinh}(\beta h)$
which holds 
for the usual effective low-energy theory of fermionic superfluids
in terms of 
{\mage ``unpaired fermion''}
quasi-particle excitations. 
As seen in Fig.~\ref{fig:m_vs_h}, the agreement is very good,  except at the largest~$h$, where interactions between quasiparticles might become significant.
The fit yields {\ora $n_{\rm qp}(T) = 0.007(8)$} for the
density of quasiparticles at ${h=0}$,
which is bounded from above by $2\,e^{-\beta E_g}$,
hence a bound on the gap
{\ora $E_g < 1.1$.}


\begin{figure}
\includegraphics[width=\columnwidth,trim=-1cm 0.5cm 2cm 0,clip]{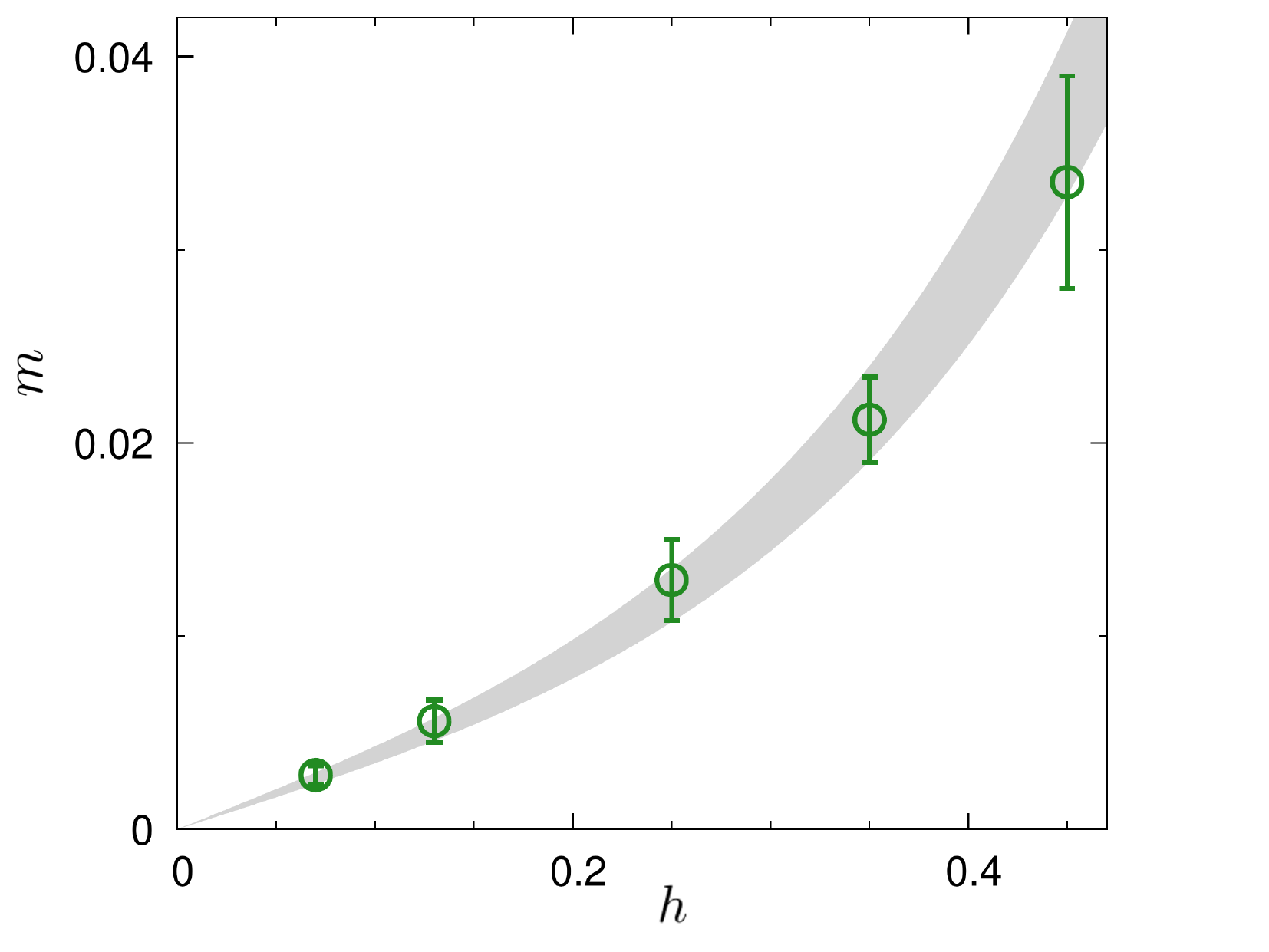}
\caption{Magnetization {\it vs.}
  Zeeman field, at
  ${T=0.19}\approx 3\,T_c/4$.
  The line is a one-parameter fit
to the low-energy theory of non-interacting quasiparticles.
  \label{fig:m_vs_h}}
          \end{figure}

Finally, we 
{\mage address} the large-order behavior of the expansion in the superconducting phase. 
We observe that $H_\xi$ contains a symmetry-breaking field
that vanishes linearly for $\xi\to1$.
This implies
that $\Or(\xi)$ contains
a 
singularity $\propto \sqrt{1-\xi}$ for $\xi\to1$~\cite{PataPokro, *BrezinWallace}.
Taylor expanding the square root gives the large-order behavior $\Or_N \propto 1/N^{3/2}$ and $P_N \propto 1/N^{5/2}$.
Hence the coefficients do not decay exponentially, as one may have naively expected, but only  like a power law.
Still, the decay is sufficiently fast for the series to be convergent.
To estimate the effect of the sub-exponential decay of the coefficients on our infinite-order extrapolations for the pressure, we supplemented the Pad\'e results with Dlog-Pad\'e
and power-law extrapolations,
and we increased the final error bars to include all obtained results. The resulting error bars are still remarkably small. In this sense, 12 loops are sufficient for accurate extrapolation.

In conclusion, 
connected diagrammatic 
expansions 
enable efficient, unbiased computations inside
superconducting phases of strongly correlated fermions.
Computations can be extended into metastable regimes,
which allows one to locate
first order transitions. 
Accurate results can be obtained even when they strongly differ from BCS mean-field predictions. 



The applicability of the approach goes far beyond the present proof of principle.
FFLO phases~\cite{FFLO}, which BCS mean-field theory to predicts to be  the true equilibrium state in a large part of the phase diagram~\cite{TormaFFLO_NJP,*TormaFFLO_3D_PRL,*TrivediFFLO,*ZhangFFLO_MF_3D}, can be accessed by
making $\Delta_0$ space dependent.}  
Stronger couplings 
{\bl may be accessible by}
using renormalized expansions,
following~\cite{RossiRDet, *SimkovicSuscept_k,SimkovicLadderCDet};
this would allow one to look for the breached-pair gapless superconducting phase~\cite{DaoPolarSF}
and to 
extend the continuous-space approach of~\cite{VanHouckeEOS, *RossiEOS, *RossiContact} to superfluid phases.
For the repulsive Hubbard model, 
the $d$-wave superconducting phase is accessible by
expanding around a momentum-dependent $\Delta_0$, as was done to second order in~\cite{metzner_perturb}.
Another natural 
extension would be to go beyond 
third-order expansions 
for open-shell nuclei~\cite{Duguet_Bogo_MPBT}
or neutron matter~\cite{UrbanNeutron_o3}.

\begin{acknowledgments}
  {\it Acknowledgements.}
  We thank   G.~Barraquand,
  G.~Biroli,   E.~Brunet, Y. Castin, N.~Dupuis, J.~Kurchan and J.~Unterberger for discussions, and E.~Burovski for 
   important help with the DDMC computations~\footnote{We used a slightly modified version of E. Burovski's original code available at 
   \href{https://github.com/ev-br/10yr_repro_challenge_35}{\sf https://github.com/ev-br/10yr\_repro\_challenge\_35}.}.
     We acknowledge support from
  the Paris \^{I}le-de-France region in the framework of DIM SIRTEQ (G.S.),
   H2020/ERC Advanced grant Critisup2 No. 743159 (F.W.),
  and the Simons Foundation through the Simons Collaboration on the
  Many Electron Problem (F.S. and M.F.).
  The Flatiron Institute is a
division of the Simons Foundation.
  This work was granted access to the HPC resources of TGCC and IDRIS under the allocations A0090510609 attributed by GENCI (Grand Equipement National de Calcul Intensif).
\end{acknowledgments}






\bibliography{felix_copy}

\end{document}